\documentclass[aps,pre,reprint,twocolumn]{revtex4-2}

\usepackage{amsmath}
\usepackage{amsfonts}
\usepackage{amssymb}
\usepackage[usenames,dvipsnames]{color}
\usepackage[colorlinks=true,linkcolor=Maroon,citecolor=Maroon,urlcolor=Maroon]{hyperref}
\usepackage{graphicx}
\usepackage{xcolor}

\usepackage[normalem]{ulem}
\definecolor{vtadd}{rgb}{0.80,0.00,0.00}
\definecolor{vtdel}{rgb}{0.00,0.00,0.80}

\newcommand{\del}[1]{\textcolor{vtdel}{\sout{#1}}}

\newcommand{\A}{\mathbf{A}}
\newcommand{\x}{\mathbf{x}}
\newcommand{\Df}{\mathbf{Df}}
\newcommand{\f}{\mathbf{f}}

\makeatletter
\def\ps@pprintTitle{%
 \let\@oddhead\@empty 
 \let\@evenhead\@empty
 \def\@oddfoot{}%
 \let\@evenfoot\@oddfoot}
 \makeatother 

\begin{document}

    \title{Non-Normal Route to Chaos}

    \author{D. Sornette$^*$}
    \author{V.R. Saiprasad}
    \author{V. Troude}
    \affiliation{
        Institute of Risk Analysis, Prediction and Management (Risks-X),
        Academy for Advanced Interdisciplinary Sciences,
        Southern University of Science and Technology, Shenzhen, China\\
        $^*$ corresponding author: dsornette@ethz.ch
           }

    \begin{abstract}
    
Deterministic chaos is usually associated with local spectral expansion:
Jacobian eigenvalues are expected to exceed unity somewhere on the
attractor. We show that this view is incomplete in dimensions \(d>1\).
For non-normal Jacobians, pointwise spectral stability can suggest
everywhere local contraction, while non-orthogonal eigenvectors still allow
transient singular-vector amplification. We construct four low-dimensional
deterministic maps realizing this mechanism: partition-reinjected,
phase-prescribed, feedback-driven, and affine-reinjected non-normal routes to
chaos. In all cases, the instantaneous Jacobian remains spectrally stable on
the attractor, with eigenvalues fixed inside the unit disk, while increasing
non-normality drives the maximal Lyapunov exponent through zero. The
positive exponent therefore describes sustained asymptotic chaos, not
transient chaos. Across the four classes, the common signature is
spectral radius \(\rho_{\mathrm{traj}}^{\max}<1\), singular value \(\sigma_{\mathrm{traj}}^{\max}>1\),
maximum Lyapunov exponent \(\lambda_1>0\), and an increase of attractor dimension. These examples
identify non-normality and recurrent reinjection of transiently amplified
directions as a deterministic route to chaos distinct from eigenvalue
instability.
    \end{abstract}

    \maketitle

\section{Introduction}

Deterministic dynamical systems
can exhibit highly irregular behavior
known as \emph{chaos}.
Since the pioneering discoveries of Lorenz \cite{Lorenz1963}, Smale \cite{Smale1967},
and others (see e.g. \cite{Ott2002,Strogatz}), chaos has been recognized as a fundamental mechanism through which simple nonlinear systems generate complex dynamics
in many natural and engineered systems.
Understanding the mechanisms that generate chaos is therefore a central problem in nonlinear science.
Classical routes to chaos, such as period-doubling cascades \cite{Feigenbaum1978,CoulletTresser1978}, intermittency \cite{MannevillePomeau}, and quasiperiodicity \cite{RuelleTakens}, have provided deep insight into how deterministic systems lose stability and develop sensitive dependence on initial conditions. Identifying the dynamical mechanisms that underlie chaotic behavior remains crucial both for theoretical understanding and for predicting or controlling complex dynamics in physical systems.

Chaos in deterministic dynamical systems $\x_{n+1}=\f(\x_n)$ with $\x_n\in\mathbb{R}^d$ is commonly diagnosed by a positive maximal Lyapunov exponent, which measures the average exponential separation of nearby trajectories and reflects sensitive dependence on initial conditions \cite{Devaney,KatokHasselblatt}. 
In one dimension, and more generally when the Jacobians $\Df(\x)$ are \emph{normal} (i.e., matrices that commute with their conjugate transpose), the growth of perturbations is tightly controlled by the instantaneous eigenvalues. 
Indeed, for normal Jacobians, the $L_2$ operator norm $\|\cdot\|_2$ (the largest singular value) equals the spectral radius, $\|\Df(\x)\|_2=\rho(\Df(\x))$.
As a result, a positive maximal Lyapunov exponent requires that the map be locally expansive along sufficiently many portions of the trajectory. In particular, if $\rho(\Df(\x))<1$ for all $\x$, then $\|\Df(\x)\|_2<1$ everywhere, implying that infinitesimal perturbations contract uniformly and that the maximal Lyapunov exponent is negative. 
This relationship naturally motivates the common identification of transitions to chaos with a form of \emph{spectral criticality}, namely the excursion of eigenvalues of the Jacobian outside the unit circle.

Here we show that identifying the instability required for chaos with
eigenvalue criticality is incomplete in dimensions \(d>1\).  Spectral
analysis captures only part of the stability landscape: the geometry of
singular values and the non-normality of the Jacobian provide additional
routes to finite-time amplification \cite{TrefethenEmbree,Schmid2007}
(the Jacobian is non-normal when it does not commute with its conjugate transpose).
In particular, chaotic dynamics can arise even when
\[
    \rho(\Df(\x_n))<1~~{\rm everywhere~along~the~attractor}~,
\]
so that no instantaneous eigenvalue instability occurs.
In this sense, chaos need not originate from an eigenvalue crossing; it can
emerge from recurrent geometric amplification, in the spirit of earlier
transient-growth scenarios in linearly stable systems
\cite{Grossmann1994}.

We demonstrate this constructively by organizing four deterministic maps
into a common taxonomy of non-normal routes to chaos. Each class is built
from the same spectrally stable non-normal core matrix, whose eigenvalues
remain fixed inside the unit disk while the eigenvector geometry is varied.
The classes differ only in the mechanism that regenerates transient
amplification: state-dependent branch selection, phase-prescribed
reorientation, endogenous feedback-driven reorientation, or affine
reinjection without modulo wrapping. Across all four cases, the same
diagnostic signature is observed:
\begin{equation}
    \rho_{\mathrm{traj}}^{\max}<1,
    \qquad
    \sigma_{\mathrm{traj}}^{\max}>1,
    \qquad
    \lambda_1>0 .
    \label{einqrttbv}
\end{equation}
Here \(\rho_{\mathrm{traj}}^{\max}\) denotes the maximum, along the
computed trajectory, of the spectral radius of the instantaneous Jacobian;
\(\sigma_{\mathrm{traj}}^{\max}\) denotes the corresponding maximum
largest singular value; and \(\lambda_1\) is the maximal Lyapunov exponent.
Thus the local eigenvalue spectrum remains contracting at every sampled
point, while the singular-value diagnostic reveals transient amplification
and the asymptotic dynamics is chaotic. 

The inequalities~(\ref{einqrttbv}) hold once the non-normality parameter
\(K\), defined below, exceeds a transition value \(K_c\). This threshold is
defined by the zero crossing of the maximal Lyapunov exponent,
\(\lambda_1(K_c)=0\). Its value depends on how each class samples,
regenerates, and reinjects the amplified directions.
This dependence is natural from the viewpoint of matrix-product dynamics:
asymptotic growth is controlled by the full operator product, not by the
instantaneous spectrum alone
\cite{Crisanti1993,troude2024illusions}. Non-normality provides the
amplifying geometry, while switching or reinjection supplies the recurrence
needed to convert transient growth into sustained Lyapunov instability.
This also connects the maps constructed below to switched and hybrid
systems, where products of individually stable operators can nevertheless
generate instability \cite{Liberzon2003,DeCarlo2000}.

This perspective also clarifies the relation to the transient-growth
scenario of Gebhardt and Grossmann \cite{Grossmann1994}. Their work already
contains the key physical intuition that non-normal transient growth, though
temporary in the linearized dynamics, can be made dynamically relevant when
the nonlinear flow continually repopulates transiently amplifying
directions. The distinction lies in how this repopulation is organized. In
\cite{Grossmann1994}, the
laminar state remains locally attracting, while finite-amplitude
perturbations or noise can trigger escape into a nonlinear chaotic regime.
In the maps studied here, by contrast, the transition is intrinsic to
autonomous bounded dynamics and is controlled directly by the non-normality
parameter \(K\). As \(K\) is increased at fixed eigenvalues, the maximal
Lyapunov exponent crosses zero even though the instantaneous spectral
radius remains below unity everywhere on the attractor. Thus, the present
route replaces the finite-amplitude trigger of the subcritical
transient-growth scenario by deterministic switching or reinjection that
recurrently restores singular-vector amplification directions.

We also distinguish the present mechanism from \emph{transient chaos} in the
sense of Lai and T\'el~\cite{LaiTel2011}, where chaotic dynamics are sustained only
over finite times on a non-attracting chaotic saddle before the orbit escapes
to a regular attractor or to infinity. In the four classes studied below, the
attractor itself is bounded and forward-invariant, no escape occurs at any
time, and the positive Lyapunov exponent characterises the asymptotic dynamics
on this attractor rather than a finite-time exit rate.

The paper is organized as follows.  Section~\ref{sec:theory} introduces the
common non-normal core matrix, the non-normality index \(K\), and the
distinction between the matrix-level transient-growth proxy and the
class-specific Lyapunov threshold \(K_c\).  Section~\ref{sec:classes}
presents the four dynamical systems: partition-reinjected,
phase-prescribed, feedback-driven, and affine-reinjected non-normal routes to
chaos.
Section \ref{sec:supp_nnr_robustness} scans the feedback strength, memory length and rotation amplitude, 
and shows that the non-normal route to chaos is found
over a broad domain of parameters in the Class III map, which is
representative of the robustness for the other classes.
Section~\ref{sec:conclusion} concludes. 
The Appendix describes the methods to calculate
the Lyapunov exponent, the spectral radius and singular values, 
the Kaplan-Yorke and box-counting, correlation fractal dimensions 
and the bifurcation projections.

\section{Spectral stability and non-normal transient amplification}
\label{sec:theory}

We first fix the notation used throughout the four classes of dynamical
systems presented below.  Consider a
discrete-time dynamical system
\[
    \x_{n+1}=\f(\x_n),
    \qquad
    \x_n\in \Omega\subset \mathbb{R}^d,
\]
where \(\Omega\) is bounded and forward invariant.  Let
\[
    J_n=\Df(\x_n)
\]
denote the Jacobian evaluated along a trajectory.  The maximal Lyapunov
exponent is
\begin{equation}
    \lambda_1(\x_0)
    =
    \limsup_{N\to\infty}
    \frac{1}{N}
    \log
    \left\|
        J_{N-1}J_{N-2}\cdots J_0
    \right\|.
    \label{eq:lyapunov_product_general}
\end{equation}
In one dimension, and more generally for normal Jacobians, Lyapunov growth is
controlled directly by instantaneous eigenvalues.  Indeed, if
\(J_nJ_n^\dagger=J_n^\dagger J_n\), then
\[
    \|J_n\|_2=\rho(J_n),
\]
so uniform spectral contraction, \(\rho(J_n)<1\), implies uniform norm
contraction and hence \(\lambda_1<0\).  In this setting, sustained chaos
requires local spectral expansion along some portion of the orbit.

For non-normal Jacobians\textcolor{brown}{,} this implication fails.  Even when all eigenvalues
remain inside the unit disk, the largest singular value
\[
    \sigma_{\max}(J_n)=\|J_n\|_2
\]
may exceed unity, allowing finite-time perturbation growth
\cite{TrefethenEmbree,Schmid2007}.  However, a single transient amplification
event is not enough to create chaos in a bounded dissipative system.  The
essential additional ingredient is recurrence: the dynamics must repeatedly
return perturbations to directions in which the next spectrally stable
Jacobian can amplify them.  The route studied below is therefore a product
mechanism.  Individual factors satisfy \(\rho(J_n)<1\), but ordered products
of non-normal and non-commuting Jacobians can have positive asymptotic growth: $\lambda_1>0$.

All four classes considered in this paper are built from the same
spectrally stable non-normal core matrix
\begin{equation}
    A_0(\kappa)
    =
    \begin{pmatrix}
        \alpha & \beta\kappa\\
        \beta/\kappa & \alpha
    \end{pmatrix},
    \qquad
    ~~\alpha,\beta > 0,~~\kappa\geq 1~ .
    \label{eq:A0_common}
\end{equation}
The eigenvalues of \(A_0(\kappa)\) are
\begin{equation}
    \lambda_{\pm}=\alpha\pm \beta,
    \label{eq:A0_eigenvalues}
\end{equation}
and are therefore independent of \(\kappa\).  Thus changing \(\kappa\) changes
the eigenvector geometry but not the local spectrum.  Unless stated
otherwise, we use
\[
    \alpha=0.7,
    \qquad
    \beta=0.2,
\]
so that
\[
    \rho(A_0)=\alpha+\beta=0.9<1.
\]
The departure from normality is measured by the common scalar index
\begin{equation}
    K
    =
    \frac{\kappa-\kappa^{-1}}{2}.
    \label{eq:K_def}
\end{equation}
This index vanishes in the normal limit \(\kappa=1\) and increases
monotonically with eigenvector non-orthogonality.

Because the same core matrix appears in all four classes, one can define a
common matrix-level proxy for the onset of transient amplification.  The
maximally alternating two-step product is obtained by applying
\(A_0(\kappa)\) and \(A_0(\kappa)^\top\), giving the symmetric positive-definite product 
$
    A_0(\kappa)A_0(\kappa)^\top .
$
The corresponding per-step growth proxy is
\begin{equation}
    \lambda_{\mathrm{proxy}}(K)
    =
    \frac{1}{2}
    \log
    \left\|
        A_0(\kappa)A_0(\kappa)^\top
    \right\|
    =
    \log \sigma_{\max}\!\left(A_0(\kappa)\right).
    \label{eq:proxy_growth}
\end{equation}
This motivates a matrix-level transient-growth threshold \(K_\sigma\), defined
by
\begin{equation}
    \sigma_{\max}\!\left(A_0(K_\sigma)\right)=1.
    \label{eq:Ksigma_def}
\end{equation}
Equivalently, using \(\lambda_\pm=\alpha\pm\beta\), this threshold can be
written as
\begin{equation}
    K_\sigma
    =
    \frac{1}{2\beta}
    \sqrt{(1-\lambda_+^2)(1-\lambda_-^2)}.
    \label{eq:Ksigma_formula}
\end{equation}
The quantity \(K_\sigma\) is common to the four classes because it depends
only on the shared matrix \(A_0(\kappa)\).  It marks the threshold value for $K$ beyond which the
local non-normal core is capable of singular-value amplification.

The actual transition to chaos, however, is not determined by
\(A_0(\kappa)\) alone.  It also depends on how the nonlinear or piecewise
dynamics samples the amplifying directions, how branch changes are generated,
and how trajectories are reinjected into a bounded set.  We therefore define
the critical value used in the figures operationally and separately for each
class. Using
$
    j\in \{\mathrm{I},\mathrm{II},\mathrm{III},\mathrm{IV}\}
$
to label the four classes of dynamical systems, then
\begin{equation}
    K_c^{(j)}
    \quad\text{is defined by}\quad
    \lambda_1^{(j)}\!\left(K_c^{(j)}\right)=0.
    \label{eq:Kc_def}
\end{equation}
Thus \(K_c^{(j)}\) is a Lyapunov transition value, not a spectral critical
point and not necessarily identical to the matrix-level proxy \(K_\sigma\).
All figures below use the normalized non-normality
$
    K/K_c^{(j)},
$
so that the observed transition for each class is aligned at
$K/K_c^{(j)}=1$.

The same diagnostics are used throughout.  Pointwise spectral stability is
measured by
\begin{equation}
    \rho_{\mathrm{traj}}^{\max}
    =
    \max_n \rho(J_n),
    \label{eq:rho_traj_max}
\end{equation}
which remains below unity in the regimes reported here.  One-step
non-normal amplification is measured by the largest singular value,
\begin{equation}
    \sigma_{\mathrm{traj}}^{\max}
    =
    \max_n \sigma_{\max}(J_n),
    \label{eq:sigma_traj_max}
\end{equation}
or by its trajectory average.  Long-time instability is measured by the
maximal Lyapunov exponent \(\lambda_1\), computed from products of Jacobians.
Finally, attractor dimensions and bifurcation projections are used as
geometric checks of the transition.  The characteristic signature of the
non-normal route is therefore provided by the inequalities (\ref{einqrttbv}),
namely Lyapunov instability and attractor complexity without instantaneous
spectral instability.

The four classes presented in Sec.~\ref{sec:classes} are not four separate
mechanisms but four realisations of the same product-growth principle, organised
by two independent design choices. The first choice is the \emph{orientation
source}, i.e.~the rule by which the orbit selects the next member of
\(\{A_0(\kappa), A_0(\kappa)^\top\}\) (or a rotated copy thereof). It can be a
state partition \(\sigma(\mathbf{x}_n)\) (Classes~I and~IV), an externally
prescribed phase \(z_n\) (Class~II), or an internally generated phase coupled
back to the planar state (Class~III). The second choice is the
\emph{boundedness mechanism}: modulo-one wrapping on the torus
(Classes~I, II, III) or branch-dependent affine translations on \(\mathbb{R}^2\)
(Class~IV). In this two-axis taxonomy, Class~IV generalizes Class~I with the
mod-one wrap replaced by affine reinjection, while Class~III extends Class~II with
the orientation phase made endogenous through feedback. Class~II is the only
class with a genuine skew-product structure, since its phase variable evolves
independently of the planar state. The class-specific Lyapunov thresholds
\(K_c^{(j)}\) introduced in Eq.~\eqref{eq:Kc_def} consequently differ across
classes only through how each class samples and recurrently visits the
amplifying directions of \(A_0(\kappa)\); the underlying matrix-level
proxy~\(K_\sigma\) (Eq.~\eqref{eq:Ksigma_formula}) is common to all four.

\section{Four classes of non-normal routes to chaos}
\label{sec:classes}

   \subsection{Mode Switching Map}

    To understand how transient amplification by non-normal stable maps 
   can lead to chaos, let us consider the minimal deterministic switching map  
    \begin{equation}
    \x_{n+1}=
    \begin{cases}
    \A \x_n \pmod{1}, & g(\x_n)>0,\\
    \A^T \x_n \pmod{1}, & g(\x_n)\le 0,
    \end{cases}
    \label{gjnhtb}
    \end{equation}
    where $\rho(\A)<1$ and $\sigma_{\max}(\A)>1$.
    The modulo operation ensures reinjection into a bounded domain.
    When alternation between $g(\x_n)>0$ and $g(\x_n)\leq0$ occurs sufficiently often,
    the leading contribution to the Lyapunov exponent is
    $\lambda \approx \frac{1}{2}\log \rho(\A^T \A) = \log \sigma_{\max}(\A)$.
    Therefore $\lambda>0$ whenever $\sigma_{\max}(\A)>1$,
    despite the spectral radius $\rho(\A)$ being smaller than $1$ at every step. This simple map (\ref{gjnhtb})  illustrates how non-normality converts
    short-time amplification into sustained exponential growth in the presence of alternation
    between the two eigendirections.
    These considerations show that the key object of interest is the maximum singular value of matrix $\A$ defined by (\ref{eq:A0_common})
    which is given by $\sigma_{\max}(\A) = \sqrt{S + \sqrt{S^2-(\alpha^2-\beta^2)^2}}$,
    where $S=\alpha^2+\beta^2+2\beta^2 K^2$.
    Expanding in powers of $K$ (\ref{eq:K_def}) gives
    \begin{equation}
    \log \sigma_{\max}(\A)
    =
    \log(\rho(\A)) +\frac{\beta}{2\alpha}K^2
    +O(K^4)~.
    \label{heynbwg}
    \end{equation}
    Expression~(\ref{heynbwg}) shows that non-normality, \(K\neq 0\), can
increase the largest singular value far above what is suggested by the
spectral radius alone. For the map~(\ref{gjnhtb}), the threshold for the
onset of chaos is determined by the condition
\(\sigma_{\max}(\A)=1\). This gives \(K_c=K_\sigma\), with
\(K_\sigma\) defined in Eq.~(\ref{eq:Ksigma_formula}).
Combining Eq.~(\ref{heynbwg}) with the approximation
$K_\sigma \simeq
\left[
    {2\alpha \over \beta}\ln (1/ \rho(\A))
\right]^{1/2}, 
$
shows that stronger spectral contraction requires stronger non-normality
for chaos to occur. In other words, the smaller the spectral radius
\(\rho(\A)\), the larger the eigenvector non-orthogonality needed for
transient amplification events to compensate the net contraction. Thus,
even when all eigenvalues are strictly contracting, increasing
non-normality alone can drive a transition from asymptotic contraction to
sustained amplification. This identifies non-normality as an independent
control parameter, distinct from eigenvalue instability.

We realize this mechanism in four deterministic classes of non-normal
routes to chaos that are presented below:  partition-reinjected, phase-prescribed, feedback-driven, and
affine-reinjected. In each class, the instantaneous Jacobian remains
pointwise spectrally stable, while the maximal Lyapunov exponent becomes
positive as the non-normality index is increased at fixed eigenvalues.

\subsection{Class I: partition-driven non-normal reorientation}
\label{sec:classI}

The first dynamical system shows that a state-dependent partition and a common
reinjection can sustain non-normal product growth even when each branch is
spectrally contracting.  The map acts on the two-dimensional torus and is
defined by
\begin{equation}
\x_{n+1}
=
A_{\sigma(\x_n)}(\kappa)\x_n+\mathbf{t}
\pmod 1,
    ~~
    \x_n=(x_n,y_n)^\top\in \mathbb{T}^2,
    \label{eq:classI_map}
\end{equation}
where the branch is selected by the state partition
\begin{equation}
    \sigma(x_n,y_n)
    =
    \begin{cases}
        +, & x_n+y_n<1,\\
        -, & x_n+y_n\geq 1.
    \end{cases}
    \label{eq:classI_partition}
\end{equation}
The two branch matrices are
\begin{equation}
    A_+(\kappa)=A_0(\kappa),
    \qquad
    A_-(\kappa)=A_0(\kappa)^\top,
    \label{eq:classI_branches}
\end{equation}
with \(A_0(\kappa)\) defined in Eq.~\eqref{eq:A0_common}.  The same
common translation \(\mathbf{t}\) is used in both branches.  In the simulations shown below
we use
\[
\mathbf{t}=(0,\tau)^\top,
\qquad
\tau=0.37.
\]
Since the translation is common to both branches, it does not affect the
Jacobian.  Its role is to maintain a nontrivial itinerary across the
partition.  Thus the local linear factors remain \(A_0(\kappa)\) and
\(A_0(\kappa)^\top\), both with spectral radius \(0.9\), while the orbit is
repeatedly reinjected across the switching boundary.

\begin{figure*}[t]
    \centering
    \includegraphics[width=0.96\textwidth]{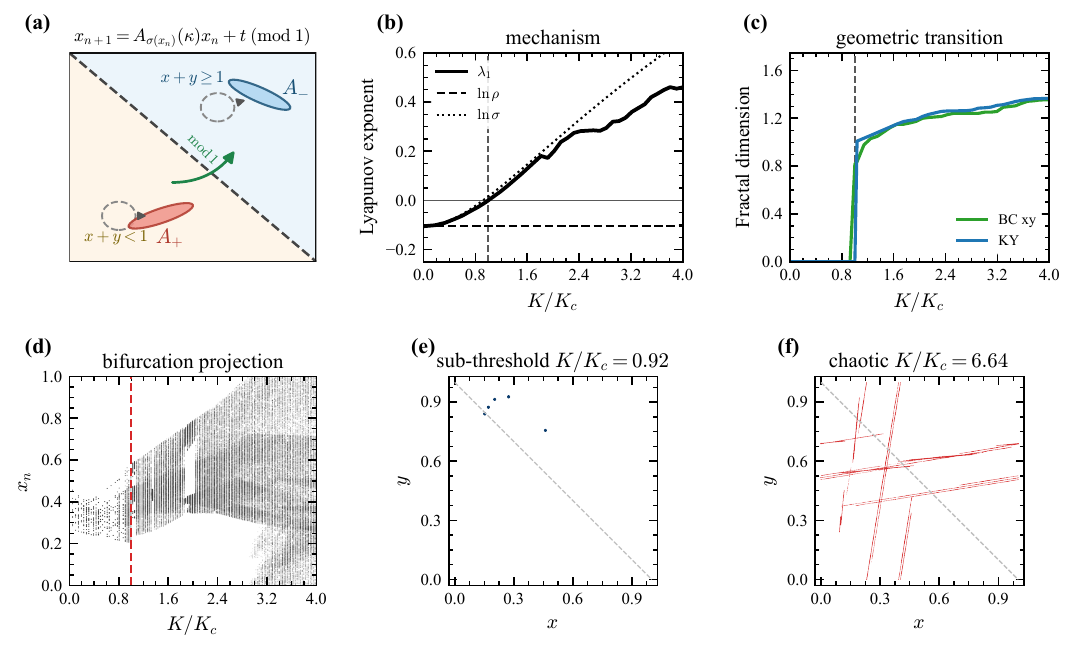}
    \caption{
   Class I: partition-driven non-normal reorientation.
    (a) Schematic of the state partition \(x+y<1\) and \(x+y\geq 1\).
    The two branches use the same spectrally stable non-normal core in
    transposed orientations,
    \(A_+(\kappa)=A_0(\kappa)\) and
    \(A_-(\kappa)=A_0(\kappa)^\top\), together with a common translation \(\mathbf{t}\).  
    The circles illustrate perturbation balls, which are mapped
into ellipses elongated along the leading local singular direction. (b) Lyapunov diagnostics as a function of the normalized non-normality
    parameter
    \(K/K_c\), where \(K_c=K_c^{(\mathrm{I})}=1.025,\) is defined by
    \(\lambda_1(K_c)=0\) for this class.  The solid curve is the maximal
    Lyapunov exponent \(\lambda_1\), the dashed curve is \(\ln\rho\), and
    the dotted curve is \(\ln\sigma\).  The dashed vertical line marks
    \(K/K_c=1\). 
    (c) Geometric diagnostics showing the box-counting estimate in the
    \((x,y)\) plane and the Kaplan--Yorke dimension. 
    (d) Bifurcation projection of \(x_n\) versus \(K/K_c\). 
    (e,f) Representative attractors below and above the transition.
    }
    \label{fig:classI_common}
\end{figure*}

Figure~\ref{fig:classI_common} shows the transition generated by this
common-reinjection mechanism.  Panel~(a) summarizes the branch geometry.
The partition determines whether the orbit is acted on by \(A_+(\kappa)\) or
by \(A_-(\kappa)\).  The common translation then returns trajectories across
the partition, so that the orbit samples both non-normal orientations.  The design of this
dynamical system separates the role of reinjection from the role of the local
Jacobian: boundedness and itinerary generation come from the translation and
the modulo operation, whereas the local tangent dynamics comes from the
spectrally stable matrices in Eq.~\eqref{eq:classI_branches}.

Panel~(b) summarises the tangent-space diagnostics of the transition.  The
trajectory-wise spectral diagnostic remains below zero in logarithmic scale,
with \(\ln\rho<0\) throughout the plotted range.  Hence the one-step
Jacobians remain pointwise spectrally stable.  The singular-value diagnostic
\(\ln\sigma\), in contrast, increases with \(K/K_c\), showing that the
available one-step transient gain grows as the eigenvectors become more
non-orthogonal.  The maximal Lyapunov exponent crosses zero at
\(K/K_c=1\), which is the operational definition of \(K_c\) for this class.
The coexistence of \(\lambda_1>0\) with \(\ln\rho<0\) identifies the
instability as a product effect rather than a local eigenvalue instability.

Panel~(c) provides a geometric check of the same transition.  Below the
Lyapunov crossing, the dimension estimates are near zero, consistent with a
finite or low-complexity attracting set.  Near \(K/K_c=1\), both the
box-counting estimate in the $(x,y)$ plane and the Kaplan--Yorke
dimension increase sharply.  For larger \(K/K_c\), the two estimates remain
above one and increase gradually, indicating that the attractor has acquired
a fractal filamentary geometric structure rather than remaining a finite periodic
set.

\begin{figure*}[t]
    \centering
    \includegraphics[width=0.96\textwidth]{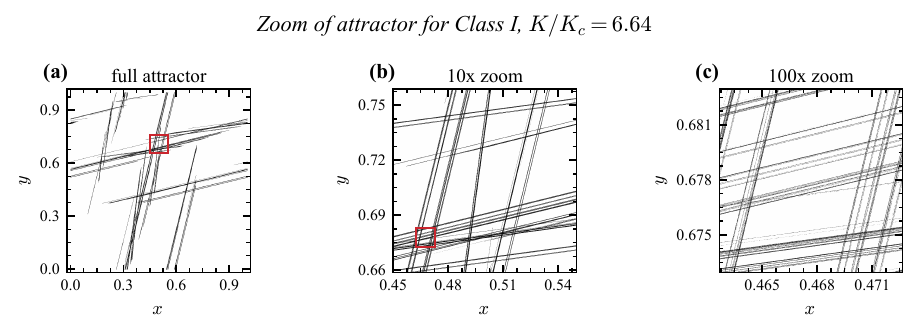}
    \caption{Successive zooms of the \((x,y)\) projection of the Class-I partition-reinjected, attractor at \(K/K_c=6.64\) shown in figure \ref{fig:classI_common}. Panels show (a) the full attractor, (b) zoom 1, and (c) zoom 2. Red rectangles indicate the region magnified in the next panel.}
    \label{fig:classI_zoom}
\end{figure*}

The bifurcation projection in panel~(d) shows how this geometric transition
appears in the observed state variable.  For \(K/K_c<1\), the recorded
values of \(x_n\) occupy a restricted set of branches.  After the Lyapunov
crossing, the projection thickens and fills a broader subset of the unit
interval.  The representative attractors in panels~(e) and (f) show the same
change in phase space.  The sub-threshold orbit at \(K/K_c=0.92\) consists of a finite set of
periodically revisited points. By contrast, the post-threshold orbit at
\(K/K_c=6.64\) forms several fractal filaments, produced by repeated
switching between the two transposed non-normal orientations
\(A_0(\kappa)\) and \(A_0(\kappa)^\top\).
Figure~\ref{fig:classI_zoom} reveals the attractor's fractal structure by
displaying several nested zooms of its filamentary geometry.

This first class establishes the minimal role of state-dependent switching
in the taxonomy.  The map does not require an independent phase or a
feedback variable.  A fixed state partition, a common translation, and the
two transposed orientations of the same non-normal core are sufficient to
produce a positive Lyapunov exponent while the one-step spectral radius
remains below unity.


A variant of Class I consists of a partition-driven
map in which reinjection is produced by sign-induced folding rather than by
an affine translation.  The map acts on the two-dimensional torus and is
defined by
\begin{equation}
    \x_{n+1}
    =
    A_{\sigma(\x_n)}(\theta)\x_n
    \pmod 1,
    ~
    \x_n=(x_n,y_n)^\top\in\mathbb{T}^2,
    \label{eq:supp_classI_sign_map}
\end{equation}
with the same partition (\ref{eq:classI_partition}).
The branch matrices are chosen as
\begin{equation}
    A_+(\theta)
    =
    R(\theta)
    \begin{pmatrix}
        a & L\\
        0 & a
    \end{pmatrix}
    R(\theta)^\top,
    \qquad
    A_-(\theta)=A_+(\theta)^\top .
    \label{eq:supp_classI_sign_branches}
\end{equation}
In the simulations we use \(a=0.5\) and \(L=1\), so that
\[
    \rho(A_\pm)=0.5<1.
\]
In expression (\ref{eq:supp_classI_sign_branches}),  \(R(\theta)\) is the planar rotation matrix of angle \(\theta\). 
There is no affine translation, $\mathbf{t}_+=\mathbf{t}_-=0$.
The parameter \(\theta\) controls the orientation of the non-normal shear
relative to the switching partition.  This variant therefore tests whether
folding produced by sign-changing matrix entries can replace the common
translation used in the design of Class I.

\begin{figure*}[t]
    \centering
    \includegraphics[width=0.96\textwidth]{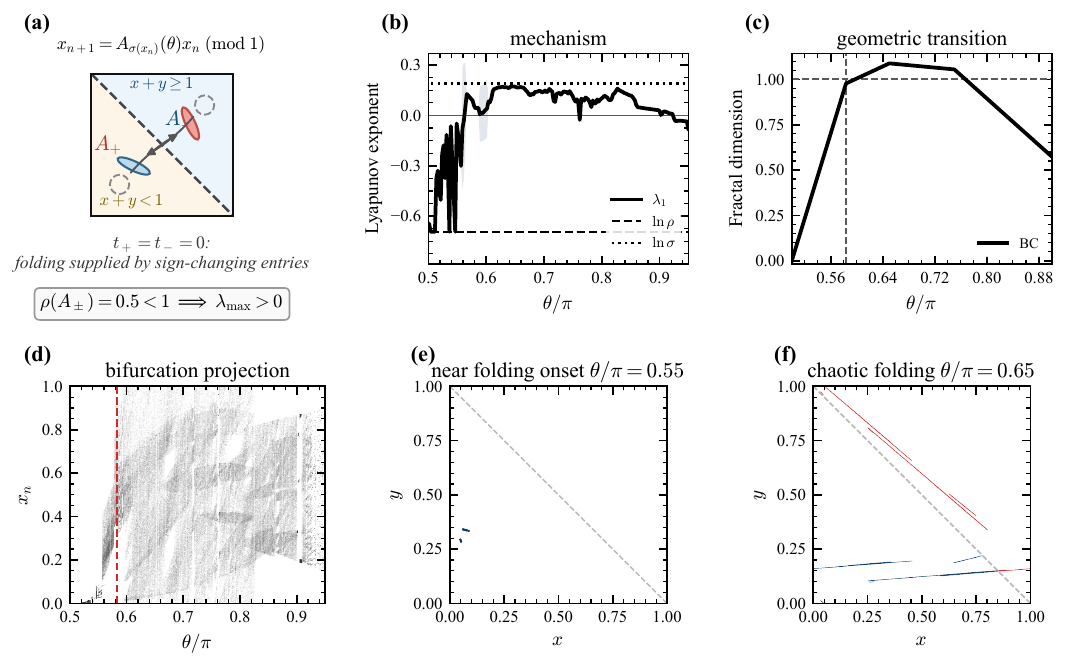}
    \caption{
     Sign-induced folding variant of Class I.
    (a) Schematic of the partition-driven map without affine translation.
    The partition \(x+y<1\) and \(x+y\geq 1\) selects between the two
    transposed branches \(A_+(\theta)\) and \(A_-(\theta)\).  
        (b) Lyapunov diagnostics as a function of the orientation parameter
    \(\theta/\pi\).  The solid curve is the maximal Lyapunov exponent
    \(\lambda_1\), the dashed curve is \(\ln\rho\) (logarithm of the spectral radius), and the dotted curve is
    \(\ln\sigma\) (logarithm of the singular value).  
    (c) Box-counting dimension estimate of the attractor. 
    (d) Bifurcation projection of \(x_n\) versus \(\theta/\pi\). 
    (e,f) Representative attractors near the folding onset and in the
    chaotic folding regime.
    }
    \label{fig:supp_classI_sign}
\end{figure*}

Here \(\theta\) is not a dynamical variable.  For each trajectory, \(\theta\)
is fixed, and the map is iterated with the two fixed matrices
\(A_+(\theta)\) and \(A_-(\theta)\).  The switching is entirely
state-dependent through Eq. (\ref{eq:classI_partition}).  The role
of \(\theta\) is only to set the static orientation of the non-normal shear
relative to the partition \(x+y=1\).  The numerical scan in
Fig.~\ref{fig:supp_classI_sign} is obtained by repeating the simulation for
fixed values of \(\theta/\pi\) over the interval shown on the horizontal
axis.  The plotted transition therefore describes how the folding mechanism
depends on the orientation of the two fixed branch matrices, not on a
time-dependent angle.

Figure~\ref{fig:supp_classI_sign} shows that the state-partitioned
mechanism can also produce positive Lyapunov exponents without an explicit
affine translation.  Panel~(a) depicts the design of the dynamical system.  The two branches are
again selected by the partition line \(x+y=1\), but the orbit is folded by the
orientation and sign structure of the matrices themselves.  Since
\(\rho(A_\pm)=0.5<1\), local spectral contraction is stronger than in the
previous Class I example.

Panel~(b) gives the tangent-space diagnostics. The singular-value diagnostic remains positive,
showing that the individual matrices can produce transient amplification
despite spectral contraction.  The maximal Lyapunov exponent becomes
positive over an interval of \(\theta/\pi\), indicating that the sequence of rotations
can visit persistently the transient amplifying directions, making the whole dynamics chaotic.

The geometric diagnostics in panels~(c) and (d) show that the positive
Lyapunov interval is accompanied by attractor thickening.  The box-counting
estimate of the fractal dimension of the attractor increases from near zero to values close to or above one, while the
bifurcation projection changes from a sparse set to a broader folded
structure.  The representative attractors in panels~(e) and (f) illustrate
the same change in phase space.  Near the onset, the orbit occupies a small
set of short segments.  In the chaotic folding regime, the attractor consists
of several folded filaments generated by alternating applications of
\(A_+(\theta)\) and \(A_-(\theta)\).

This example shows that the partition-driven
mechanism is not tied to the translation structure used in the previous example (\ref{eq:supp_classI_sign_map}).
The essential ingredients remain the presence of spectrally
stable non-normal branches, state-dependent switching, and bounded
reinjection of trajectories into amplifying directions.

\subsection{Class II: phase-prescribed non-normal reorientation}
\label{sec:classII}

The second dynamical system replaces state-partitioned switching by an
externally prescribed modulo-one phase drive. The planar state is advanced by
a non-normal operator whose orientation is controlled by an independent
phase variable \(z_n\):
\begin{equation}
   \mathbf{u}_{n+1}
=
A(z_n;\kappa)\mathbf{u}_n
\pmod 1,
~
\mathbf{u}_n=(x_n,y_n)^\top\in\mathbb{T}^2,
    \label{eq:classII_u}
\end{equation}
with
\begin{equation}
    z_{n+1}
    =
    \alpha_z z_n+\gamma
    \pmod 1.
    \label{eq:classII_z}
\end{equation}
The matrix applied to the planar variables is
\begin{equation}
    A(z;\kappa)
    =
    R(\Omega z)A_0(\kappa)R(\Omega z)^\top,
    \label{eq:classII_Az}
\end{equation}
where \(A_0(\kappa)\) is the common non-normal core 
\eqref{eq:A0_common}.  In the simulations we use
\[
    \alpha_z=\frac{2}{3},
    \qquad
    \gamma=\frac{3}{4},
    \qquad
    \Omega=2\pi.
\]
The \(z\)-dynamics does not depend on \(\mathbf{u}_n\).  It therefore generates a
prescribed sequence of rotations driven by $R(\Omega z)$, rather than a state-dependent or
feedback-generated sequence.  Since \(A(z;\kappa)\) is orthogonally similar
to \(A_0(\kappa)\), its planar eigenvalues remain
\(\alpha\pm\beta\) for all \(z\) and all \(\kappa\).  The full skew-product
Jacobian contains a coupling column from \(z_n\) to \(\mathbf{u}_{n+1}\), but its
pointwise spectral radius remains below unity over the reported range.

\begin{figure*}[t]
    \centering
    \includegraphics[width=0.96\textwidth]{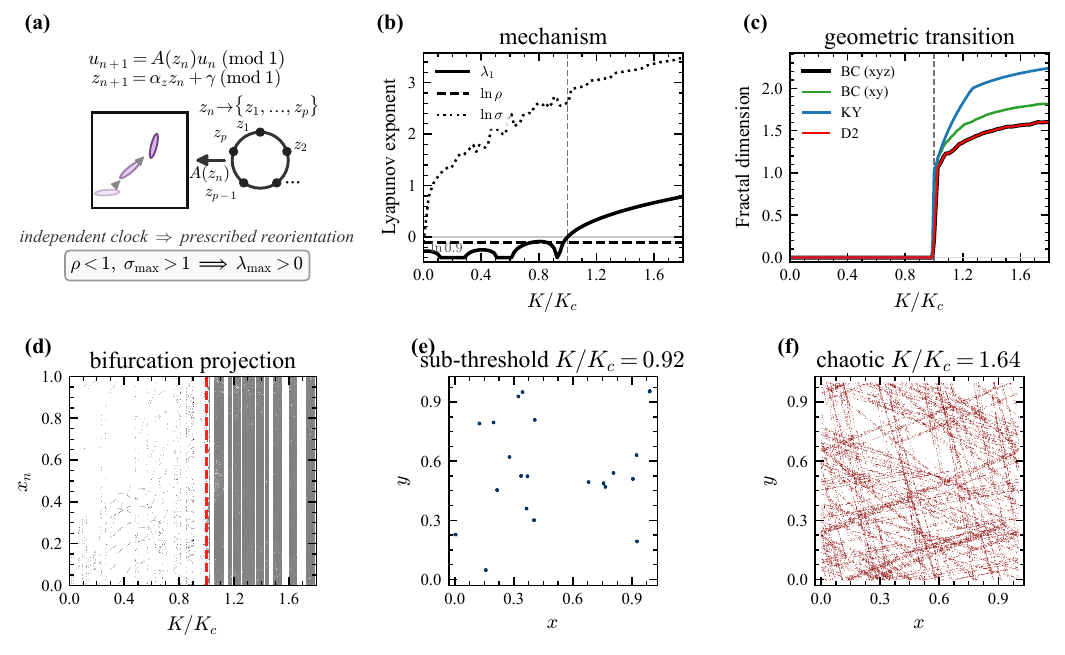}
    \caption{
    Class II: phase-prescribed non-normal reorientation.
    (a) Schematic of the skew-product map.  The phase variable \(z_n\)
    evolves independently and prescribes the orientation of the non-normal
    core \(A_0(\kappa)\). 
    (b) Lyapunov diagnostics as a function of the normalized non-normality
    parameter \(K/K_c\), where \(K_c=K_c^{(\mathrm{II})}=~5.07\) is defined
    by \(\lambda_1(K_c)=0\) for this class.  The solid curve is the maximal
    Lyapunov exponent \(\lambda_1\), the dashed curve is \(\ln\rho\) (logarithm of the largest spectral radius of the Jacobian
    along the attractor), and
    the dotted curve is \(\ln\sigma\) (logarithm of maximal singular value). 
    (c) Attractor-dimension diagnostics: box-counting estimates in
    \((x,y,z)\) and in the \((x,y)\) projection, the Kaplan--Yorke
    dimension, and the correlation-dimension estimate \(D_2\). 
    (d) Bifurcation projection of \(x_n\) versus \(K/K_c\). 
    (e,f) Representative \((x,y)\) projections below and above the
    transition.
    }
    \label{fig:classII_clock}
\end{figure*}

Figure~\ref{fig:classII_clock} shows that the modulo-one phase drive is sufficient
to convert transient non-normal gain into a sustained chaotic attractor with positive Lyapunov exponent.
Panel~(a) illustrates the two parts of the skew product.  The planar state
\(\mathbf{u}_n\) is multiplied by \(A(z_n;\kappa)\), while the phase variable follows
Eq.~\eqref{eq:classII_z}.  The itinerary of matrix orientations is therefore
determined by the phase dynamics rather than selected by the current position of
the planar state.  This distinguishes Class II from the state-partitioned
mechanism of Class I.

Panel~(b) summarises the tangent-space diagnostics of
the transition.  The dashed curve
\(\ln\rho\) remains negative, indicating pointwise spectral stability of the
one-step Jacobian over all points of the attractor.  \(\ln\sigma\) grows with \(K/K_c\),
showing that the same fixed eigenvalues support increasing one-step
transient gain quantified by the maximal singular value as the core matrix becomes more non-normal.  The maximal
Lyapunov exponent crosses zero at \(K/K_c=1\).  The transition 
occurs while the local spectrum remains subcritical, and the growth results
from ordered products of reoriented non-normal matrices, rather than
from local eigenvalue crossings.

The geometric diagnostics in panel~(c) show the corresponding change in the
attractor.  Below the transition the dimension estimates remain near zero,
consistent with a finite or low-complexity orbit under the prescribed phase dynamics.
At the Lyapunov zero-crossing, the full-space box-counting estimate, the planar
projection estimate, the Kaplan--Yorke dimension, and the correlation
dimension all increase to values larger than $1$.  Above the transition, the full-space and
Kaplan--Yorke estimates continue to increase, while the planar estimates
remain lower, indicating that the phase coordinate contributes to the
sampled attractor geometry.

\begin{figure*}[t]
    \centering
    \includegraphics[width=0.96\textwidth]{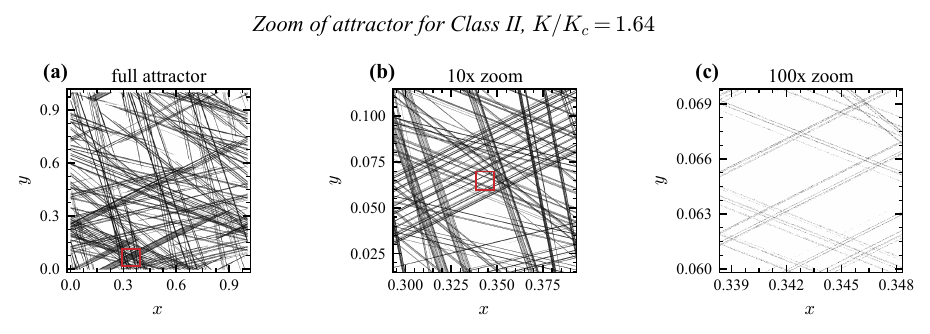}
    \caption{Successive zooms of the \((x,y)\) projection of the Class-II phase-prescribed attractor at \(K/K_c=1.64\) shown in figure \ref{fig:classII_clock}. 
    Panels show (a) the full attractor, (b) zoom 1, and (c) zoom 2. Red rectangles indicate the region magnified in the next panel.}
    \label{fig:classII_zoom}
\end{figure*}

Panel~(d) shows the evolution of the attractor projection along $x_n$ and 
its drastic change at the transition.
For \(K/K_c<1\), the projected attractor occupies a sparse set of
branches.  For \(K/K_c>1\), the projection becomes dense across much of the
unit interval.  Panels~(e) and (f) are representative of the shape 
of the attractors in the $(x,y)$ plane below and above the transition: 
the sub-threshold orbit at \(K/K_c=0.92\) is
a sparse set of points, whereas the post-threshold orbit at \(K/K_c=1.64\)
forms a dense fractal filamentary cloud in the \((x,y)\) plane.
Figure~\ref{fig:classII_zoom} reveals the attractor's fractal structure by
displaying several nested zooms of its filamentary geometry.

Class II isolates the role of prescribed reorientation. The planar state is
advanced through a deterministic sequence of orientations, rather than
through state-dependent branch selection or feedback from the amplified
variables. The phase dynamics repeatedly exposes the state to transiently
amplifying directions. This shows that non-normal product growth can be
organized by an imposed deterministic itinerary while maintaining pointwise
spectral stability.

\subsection{Class III: feedback-driven non-normal reorientation}
\label{sec:classIII}

\begin{figure*}[t]
    \centering
    \includegraphics[width=0.96\textwidth]{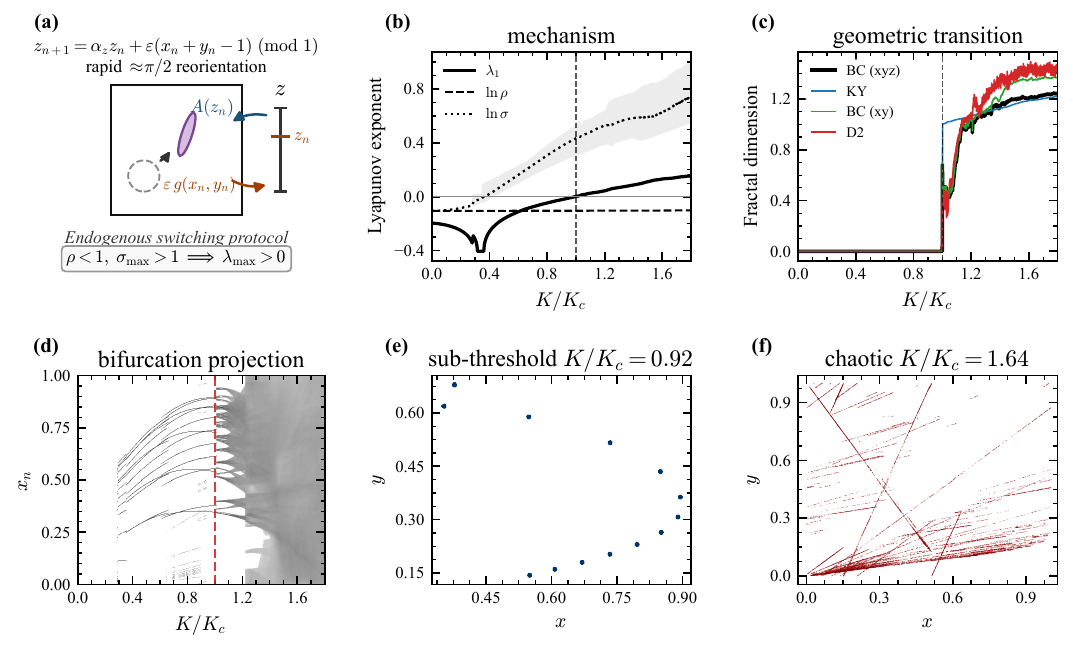}
    \caption{
    Class III: feedback-driven non-normal reorientation.
    (a) Schematic of the dynamics of the 3D map (\ref{eq:classIII_u})-(\ref{eq:classIII_z}).  
    The planar state
    \(\mathbf{u}_n=(x_n,y_n)^\top\) is acted on by the rotated non-normal
    matrix \(A(z_n;\kappa)\) (\ref{eq:classII_Az}), which depends on the phase \(z_n\)  (\ref{eq:classIII_z}),
    whose dynamics depend on the planar state via the term \(\varepsilon g(x_n,y_n)\). 
    (b) Lyapunov diagnostics as a function of the normalized non-normality
    parameter \(K/K_c\), where \(K_c=K_c^{(\mathrm{III})}=~1.953\) is the value of $K$ for which the
    Lyapunov exponent crosses zero for this class.  The solid curve is \(\lambda_1\), the
    dashed curve is \(\ln\rho\), and the dotted curve is \(\ln\sigma\);
    the shaded band quantifies the variability of the singular values along the orbit. 
    (c) Attractor-dimension diagnostics, including box-counting estimates in
    \((x,y,z)\) and in the \((x,y)\) projection, the Kaplan--Yorke
    dimension, and the correlation-dimension estimate \(D_2\). 
    (d) Bifurcation projection of \(x_n\) versus \(K/K_c\). 
    (e,f) Representative \((x,y)\) projections below and above the
    transition.
    }
    \label{fig:classIII_nnsrt}
\end{figure*}

The third dynamical system makes the reorientation dynamics endogenous.  The
planar state is again advanced by a rotated copy of the common non-normal
core, but the phase variable controlling the orientation is also influenced by the
$\mathbf{u}_{n}$ state itself:
\begin{equation}
    \mathbf{u}_{n+1}
    =
    A(z_n;\kappa)\mathbf{u}_n
    \pmod 1,
   ~
    \mathbf{u}_n=(x_n,y_n)^\top\in\mathbb{T}^2,
    \label{eq:classIII_u}
\end{equation}
\begin{equation}
    z_{n+1}
    =
    \alpha_z z_n+\varepsilon g(\mathbf{u}_n)
    \pmod 1.
    \label{eq:classIII_z}
\end{equation}
We use again $A(z;\kappa)$ given by expression (\ref{eq:classII_Az})
with \(A_0(\kappa)\) defined in Eq.~\eqref{eq:A0_common}.  The simulations
shown in Fig.~\ref{fig:classIII_nnsrt} use
\[
    \alpha_z=0.5,
    ~~
    \varepsilon=10^{-3},
   ~~
    \Omega=\pi/2,
   ~~
    g(x,y)=x+y-1.
\]

Figure~\ref{fig:classIII_nnsrt} (a) summarizes the feedback loop
embodied in (\ref{eq:classIII_u})-(\ref{eq:classIII_z}).
The current value of
\(z_n\) determines the orientation of the non-normal operator acting on
\((x_n,y_n)\).  The updated planar state then feeds back into the next value
of \(z_n\) through \(\varepsilon g(x_n,y_n)\). Differently from class-II maps, the orbit does not merely follow a prescribed sequence of
orientations, but participates in generating that sequence.

Panel~(b) gives the tangent-space diagnostics.  
\(\ln\rho\) remains negative throughout the transition, so the instantaneous
Jacobian remains pointwise spectrally stable along the sampled attractor.
\(\ln\sigma\) increases with \(K/K_c\), with a visible
trajectory-to-trajectory or orbitwise variability band, indicating increasing
availability of transient singular-vector amplification.  
Thus, for \(K/K_c>1\), the positive Lyapunov exponent does not arise from
a local spectral-radius crossing. It arises instead from products of
feedback-generated non-normal Jacobians, whose changing singular directions
recurrently convert transient amplification into sustained asymptotic
growth.

Panel~(c) characterised the attractor geometry via fractal dimensions. Their
estimates remain near zero below the transition.  Near \(K/K_c=1\), the dimension estimates leave the periodic regime and
increase across the chaotic range. The full-space box-counting estimate,
the projected box-counting estimate, the Kaplan--Yorke dimension, and the
correlation dimension need not coincide quantitatively, as expected for
finite-sample estimators applied to different geometric projections or
measures. Nevertheless, they consistently indicate the same qualitative
transition: from a low-complexity periodic set to a higher-dimensional
strange attractor.

\begin{figure*}[t]
    \centering
    \includegraphics[width=0.96\textwidth]{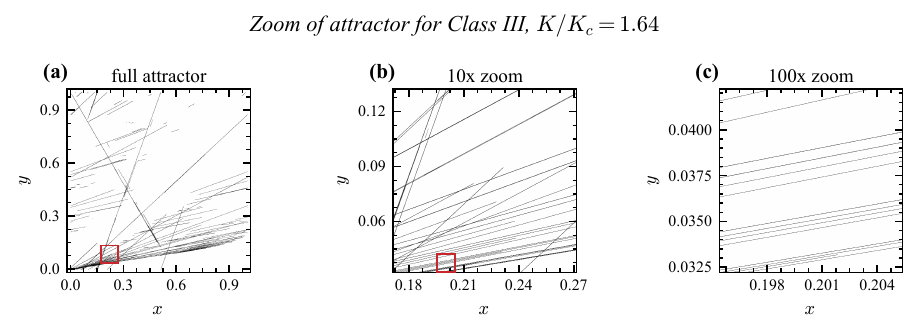}
    \caption{Successive zooms of the \((x,y)\) projection of the Class-III feedback-driven, attractor at \(K/K_c= 1.64\) shown in figure \ref{fig:classIII_nnsrt}. 
    Panels show (a) the full attractor, (b) zoom 1, and (c) zoom 2. Red rectangles indicate the region magnified in the next panel.}
    \label{fig:classIII_zoom}
\end{figure*}

Panel~(d) shows how the projection of the attractor onto the \(x_n\)
coordinate evolves with \(K/K_c\), revealing a sharp structural
reorganization at the Lyapunov crossing. Before the transition, the
projection is supported on a small number of well-organized branches,
consistent with regular or periodic dynamics. After the crossing, the
projection develops a denser multibranch structure over a wider range of
\(x_n\) values, indicating the onset of sustained chaotic dynamics. The
phase-space projections in panels~(e) and~(f) show the same reorganization
geometrically. At \(K/K_c=0.92\), the orbit remains confined to a small
structured set. At \(K/K_c=1.64\), it develops many reinjected filaments,
consistent with a strange attractor generated by repeated state-driven
reorientation. Figure~\ref{fig:classIII_zoom} reveals the attractor's fractal structure by
displaying several nested zooms of its filamentary geometry.

Class III differs from Class II in how the successive orientations of the
non-normal operator are generated. In
Class II, the sequence is prescribed by an independent mod-one phase dynamics.  In Class III,
it is generated endogenously by feedback from the state itself.
The same non-normal core matrix is used in both cases, but the regeneration
of transient growth is self-organized rather than externally prescribed.

Class III realizes the product-growth mechanism of
Sec.~\ref{sec:theory} through feedback. When \(\varepsilon=0\), the
auxiliary variable \(z_n\) relaxes to a fixed point. The orientation of
\(A(z_n;\kappa)\) then becomes fixed, so the planar dynamics is governed by
a single spectrally stable non-normal operator. Transient amplification may
occur over short times, but it is not regenerated along the orbit; the
asymptotic Lyapunov exponent therefore remains negative.
When \(\varepsilon\neq 0\), the feedback term
\(\varepsilon g(x_n,y_n)\) prevents \(z_n\) from settling to a fixed value.
Instead, the planar state continually perturbs the orientation variable,
producing a state-dependent sequence of non-normal matrices
\(A(z_n;\kappa)\). These reorientations repeatedly place the orbit near
directions of large singular-vector gain, thereby converting transient
non-normal amplification into sustained product growth. The modulo-one
reinjection keeps the dynamics bounded, while the feedback through \(z_n\)
regenerates the amplifying orientations needed for a positive Lyapunov
exponent.

\subsection{Class IV: affine-reinjected non-normal reorientation}
\label{sec:classIV}

\begin{figure*}[t]
    \centering
    \includegraphics[width=0.96\textwidth]{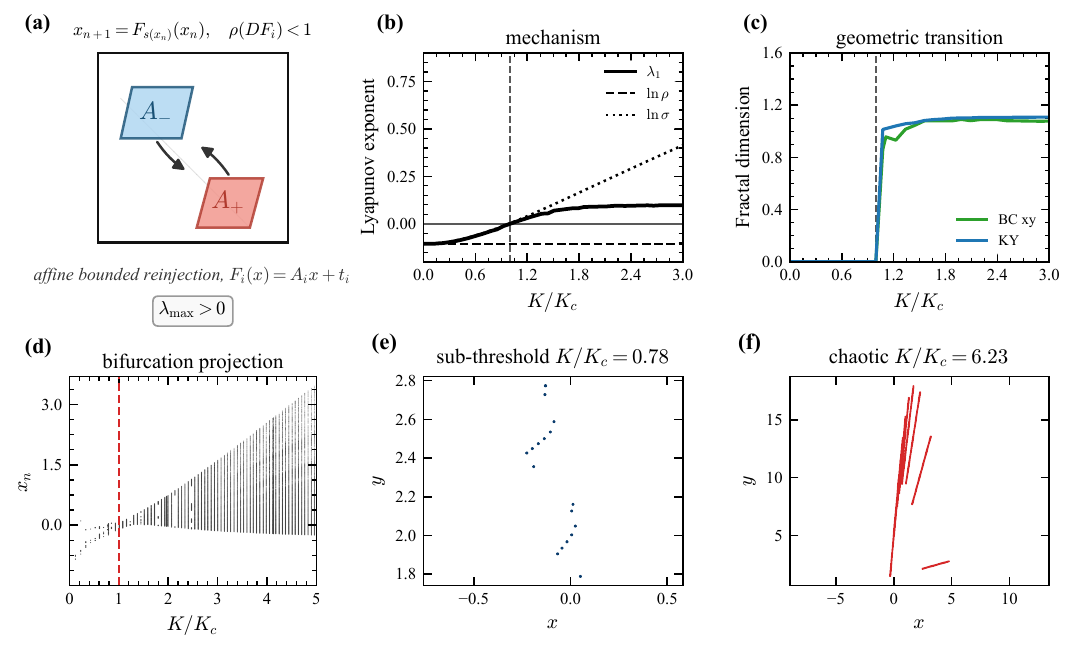}
    \caption{
   Class IV: affine-reinjected non-normal reorientation.
   (a) Schematic of the piecewise affine map.  The two branches use the
    same spectrally stable non-normal core (\ref{eq:A0_common}) in transposed orientations,
    \(A_+(\kappa)=A_0(\kappa)\) and
    \(A_-(\kappa)=A_0(\kappa)^\top\), while branch-dependent translations
    provide bounded reinjection in \(\mathbb{R}^2\). 
    (b) Lyapunov diagnostics as a function of the normalized non-normality
    parameter \(K/K_c\), where \(K_c=K_c^{(\mathrm{IV})}=~0.956\) is defined by
    the Lyapunov crossing for this class.  The solid curve is
    \(\lambda_1\), the dashed curve is \(\ln\rho\), and the dotted curve is
    \(\ln\sigma\). 
    (c) Geometric diagnostics showing the planar box-counting estimate and
    the Kaplan--Yorke dimension. 
    (d) Bifurcation projection of \(x_n\) versus \(K/K_c\). 
    (e,f) Representative attractors below and above the transition.
    }
    \label{fig:classIV_affine}
\end{figure*}

The fourth dynamical system shows that modulo wrapping is not required
for the non-normal route to chaos. 
Boundedness is instead
provided by affine reinjection through branch-dependent translations.  The
map acts on \(\mathbf{u}_n=(x_n,y_n)^\top\in\mathbb{R}^2\) and is defined by
\begin{equation}
    \mathbf{u}_{n+1}
    =
    A_{\sigma(\mathbf{u}_n)}(\kappa)\mathbf{u}_n
    +
    \mathbf{t}_{\sigma(\mathbf{u}_n)},
    \label{eq:classIV_map}
\end{equation}
where the branch is selected by an affine partition,
\begin{equation}
    \sigma(\mathbf{u}_n)
    =
    \begin{cases}
        +, & h(\mathbf{u}_n)<0,\\
        -, & h(\mathbf{u}_n)\geq 0,
    \end{cases}
    \qquad
    h(\mathbf{u})=h_0x+h_1y-h_c .
    \label{eq:classIV_partition}
\end{equation}
The linear parts use the same reciprocal non-normal matrix (\ref{eq:A0_common}) as in the preceding
classes,
\begin{equation}
    A_+(\kappa)=A_0(\kappa),
    \qquad
    A_-(\kappa)=A_0(\kappa)^\top .
    \label{eq:classIV_branches}
\end{equation}
The branch translations \(\mathbf{t}_+\) and \(\mathbf{t}_-\) provide
reinjection into a bounded region of the plane.  In the numerical
implementation used for Fig.~\ref{fig:classIV_affine}, the partition and
translations ensure that the orbit remains bounded over the reported
parameter range while the sampled Jacobians remain spectrally stable.

Figure~\ref{fig:classIV_affine} shows the same analysis as for the previous three classes
for parameters $h_0=0.828, h_1=0.561,\ h_c=1.198$
and translation vectors $t_+=(-0.942,1.097),  t_-=(-0.175,0.274)$.
Panel~(a) shows the geometric structure of the map. Each branch is affine: the linear factor 
is spectrally stable yet non-normal, and the branch-dependent translation is distinct for each branch. 
These translations serve the same reinjection role as the modulo-one folding 
in the previous classes, namely they redirect the orbit into the domain where the non-normal factor 
of the next branch takes effect, while leaving the one-step Jacobian equal to one of the two matrices in Eq.~\eqref{eq:classIV_branches}.

Panel~(b) shows the tangent-space diagnostics. The spectral-radius curve is everywhere negative 
(plotted on a logarithmic scale), confirming that the sampled one-step Jacobians remain 
spectrally contracting throughout. The singular-value diagnostic grows with \(K/K_c\):
the eigenvalues are fixed, yet the increasing non-normality of the core matrix 
amplifies transient growth. The maximal Lyapunov exponent crosses zero at the 
class-specific threshold \(K/K_c=1\) with $K_c=0.956$.
The transition is therefore driven by non-normal transient amplification alone,
while no local spectral instability occurs, and no modulo operation is involved.

Panel~(c) characterises the attractor geometry through fractal dimension estimates. 
Both the box-counting dimension $D_0$ and the Kaplan-Yorke dimension $D_{KY}$
rise near the Lyapunov crossing and remain above unity in the chaotic regime. 
For a two-dimensional map, both quantities are expected to lie in $(1,2)$, as is appropriate 
for a filamentary strange attractor embedded in the plane. Their mutual agreement 
across the plotted range supports the conclusion that the post-threshold invariant 
set has genuine fractal structure and that it is not a finite union of affine images.

\begin{figure*}[t]
    \centering
    \includegraphics[width=0.96\textwidth]{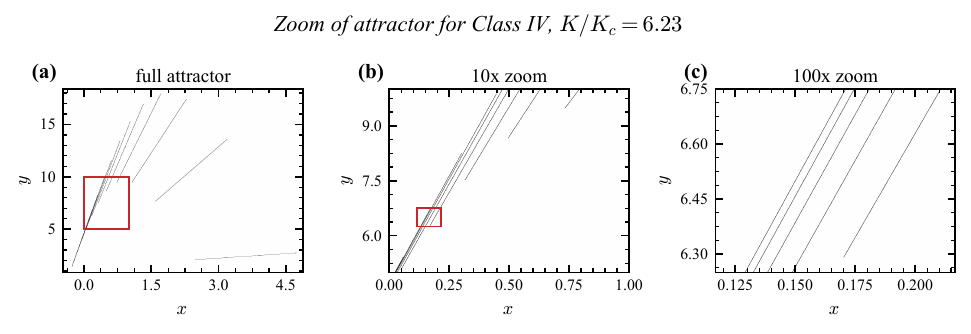}
    \caption{Successive zooms of the \((x,y)\) projection of the Class-IV affine-reinjected attractor at \textcolor{brown}{\(K/K_c=6.23\)} shown in figure \ref{fig:classIV_affine}.
    Panels show (a) the full attractor, (b) zoom 1, and (c) zoom 2. Red rectangles indicate the region magnified in the next panel.}
    \label{fig:classIV_zoom}
\end{figure*}

The bifurcation projection in panel~(d) shows how the distribution of
recorded \(x_n\) values changes across the transition. Below \(K/K_c=1\),
the projection is confined to a narrow set of organized branches. Above the
transition, it occupies a wider range of \(x_n\) values and resolves into
many thin affine strands, indicating repeated stretching and reinjection.
The representative attractors in panels~(e) and~(f) show the same
reorganization in phase space. The sub-threshold case remains supported on
a small collection of affine segments, whereas the post-threshold case
contains several stretched and reinjected filaments generated by alternating
applications of \(A_0(\kappa)\) and \(A_0(\kappa)^\top\).
Figure~\ref{fig:classIV_zoom} reveals the attractor's fractal structure by
displaying several nested zooms of its filamentary geometry.

Class IV completes the taxonomy by separating non-normal chaos from
modulo-one boundedness. The local matrices remain spectrally stable and are
drawn from the same non-normal core (\ref{eq:A0_common}), but recurrence is produced by affine
reinjection rather than torus wrapping. This shows that the mechanism
identified in the preceding classes is not an artifact of the modulo-one
operation. The essential ingredients are bounded recurrence and repeated
repopulation of transiently amplified directions, not a specific mechanism
for enforcing boundedness.


\section{Parameter robustness of non-normal routes to chaos} \label{sec:supp_nnr_robustness} 

The four classes introduced above demonstrate the same mechanism in distinct dynamical settings: a transition to chaos can be induced by increasing non-normality at fixed eigenvalues, while the instantaneous Jacobian remains pointwise spectrally stable on the attractor. In the partition-reinjected, phase-prescribed, feedback-driven, and affine-reinjected non-normal routes, the local spectral radius stays below unity, so each sampled one-step Jacobian appears contracting from its eigenvalues alone. Nevertheless, recurrent repopulation of transiently amplifying directions allows products of non-normal Jacobians to generate a positive maximal Lyapunov exponent.

We now test whether this mechanism is robust to changes in auxiliary parameters, rather than tied to a special numerical choice. We focus on the feedback-driven class as a representative case because the non-normal operator is reoriented by an endogenous variable \(z_n\), whose dynamics are controlled by the feedback amplitude \(\varepsilon\), the memory parameter \(\alpha_z\), and the rotation scale \(\Omega\). These parameters determine how efficiently transiently amplified directions are regenerated, while the non-normality index \(K\) controls the strength of the singular-vector amplification at fixed eigenvalues.

The robustness tests are organized around the three ingredients of the repeated-amplification mechanism. The feedback amplitude \(\varepsilon\) sets the strength with which the planar state perturbs the internal orientation variable. The memory parameter \(\alpha_z\) controls the persistence of this variable from one iterate to the next. The rotation scale \(\Omega\) determines how strongly changes in \(z_n\) reorient the non-normal amplifying directions. For each sweep, we vary one of these parameters together with the non-normality index \(K\), and identify the clean non-normal chaotic regime by the joint criterion 
\begin{equation} 
\lambda_1>0, \qquad \rho_{\max}<1, 
\label{eq:clean_nnr_criterion} 
\end{equation}
where \(\rho_{\max}\) denotes the maximal pointwise spectral radius of the Jacobian along the sampled trajectory (\(\rho_{\max}\equiv\rho_{\mathrm{traj}}^{\max}\) of Eq.~\eqref{einqrttbv}; the shorter notation is used in this section for legibility). This criterion excludes conventional spectral-expansion routes to chaos and isolates the regime in which asymptotic Lyapunov instability coexists with pointwise spectral contraction.

\begin{figure*}
\centering
\includegraphics[width=\textwidth]{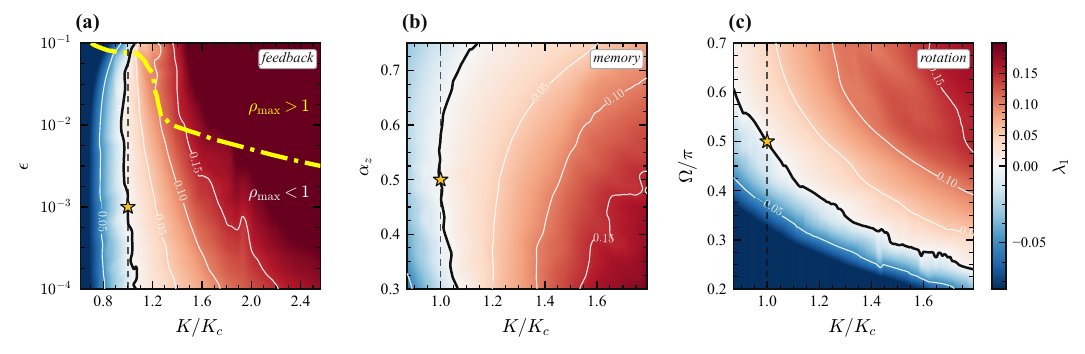}
\caption{Colors show the maximal Lyapunov exponent \(\lambda_1\) in three two-parameter sweeps for the feedback-driven Class III route: (a) feedback amplitude \(\varepsilon\), (b) memory parameter \(\alpha_z\), and (c) rotation scale \(\Omega/\pi\), each plotted against the normalized non-normality strength \(K/K_c\). Here \(K_c=1.953\) is the Lyapunov crossing for the baseline parameter set \((\alpha_z,\varepsilon,\Omega)=(0.5,10^{-3},\pi/2)\), shown by the star. White contours indicate representative values of \(\lambda_1\). The black contour marks \(\lambda_1=0\), and the dashed vertical line marks \(K/K_c=1\). In panel~(a), the yellow dash-dotted contour marks \(\rho_{\max}=1\), separating the pure non-normal region \(\rho_{\max}<1\) from the region where local spectral expansion contributes. For panels~(b) and~(c), the plotted parameter ranges satisfy \(\rho_{\max}<1\) throughout.}
\label{fig:parametric_robustness}
\end{figure*}

Figure~\ref{fig:parametric_robustness} shows that the non-normal chaotic regime persists over finite intervals of all three auxiliary parameters. 
Panel~(a) shows the dependence of the maximal Lyapunov exponent in the plane $(K/K_c, \epsilon)$.
The thin black contour marks the stability boundary
\(\lambda_1=0\), and the thick yellow dash-dotted contour marks the
boundary \(\rho_{\max}=1\). Here, \(\rho_{\max}\) denotes the maximum,
along the sampled trajectory, of the pointwise spectral radius of the full
Jacobian.
The region with \(\lambda_1>0\) extends up to values of \(\epsilon\) close
to \(0.1\). However, the simultaneous conditions \(\lambda_1>0\) and
\(\rho_{\max}<1\) are satisfied only in the subregion \(K/K_c<1.2\). This
subregion therefore corresponds to the clean non-normal route to chaos: the
dynamics is chaotic even though the instantaneous Jacobian is locally
spectrally contracting everywhere along the sampled trajectory.
Beyond the \(\rho_{\max}=1\) boundary, local spectral expansion is also
present and can contribute to the instability. The dynamics may therefore
remain chaotic, but it should no longer be classified as arising from a
purely non-normal mechanism.

Panel~(b) shows that the maximal Lyapunov exponent remains positive for intermediate memory, with an onset threshold close to the baseline value for \(\alpha_z\simeq 0.45\)--\(0.6\). This is consistent with the role of \(z_n\) as an endogenous orientation variable: when \(\alpha_z\) is too small, \(z_n\) relaxes too rapidly; when \(\alpha_z\) is too large, it retains too much of its previous value and reorientation becomes less effective. Panel~(c) shows that the transition is not tied to the baseline choice \(\Omega=\pi/2\). Increasing the rotation scale generally lowers the required non-normality threshold, as expected when recurrent reorientation of transiently amplifying directions is the mechanism sustaining chaotic growth.

These sweeps support the robustness of the non-normal route. The primary control parameter remains the non-normality index \(K\): increasing \(K\) strengthens transient singular-vector amplification while the eigenvalues remain fixed inside the unit disk. The auxiliary parameters do not create a different mechanism; instead, they modulate how efficiently the transient gain is regenerated. The feedback amplitude \(\varepsilon\) must be large enough to drive recurrent changes in \(z_n\), but not so large that the clean spectrally subcritical regime is lost. The memory parameter \(\alpha_z\) controls the time scale over which the orientation variable retains its past state. The rotation scale \(\Omega\) controls how strongly changes in \(z_n\) reorient the amplifying directions. These results show that the transition occupies a large open region of parameter space.

To complement the two-parameter sweeps in Fig.~\ref{fig:parametric_robustness}, we also scan the three-dimensional parameter volume \((K,\alpha_z,\varepsilon)\). 
Since this volume cannot be displayed directly in a single two-dimensional plot, we summarize the \(K\)-direction in two complementary ways. 
First, for each pair \((\alpha_z,\varepsilon)\), we define the clean non-normal onset threshold 
\begin{equation} 
\begin{aligned} 
K_c^{\rm NNR}(\alpha_z,\varepsilon) = \min \Bigl\{ K:\;& \lambda_1(K,\alpha_z,\varepsilon)>0,\\ 
& \rho_{\max}(K,\alpha_z,\varepsilon)<1 \Bigr\}. 
\end{aligned} 
\label{eq:Kc_NNR_robust} 
\end{equation} 
Second, we define the fraction of scanned \(K\)-values satisfying the same clean non-normal-chaos criterion, 
\begin{equation} 
\begin{aligned} 
P_{\rm NNR}(\alpha_z,\varepsilon) = \frac{1}{N_K} \sum_K \mathbf{1} \Bigl\{ 
& \lambda_1(K,\alpha_z,\varepsilon)>0,\\ &\rho_{\max}(K,\alpha_z,\varepsilon)<1 \Bigr\}. 
\end{aligned} 
\label{eq:P_NNR} 
\end{equation} 
The threshold \(K_c^{\rm NNR}\) measures how much non-normality is required for onset, whereas \(P_{\rm NNR}\) measures the width of the clean chaotic regime along the scanned \(K\)-direction.

\begin{figure*} 
\centering \includegraphics[width=0.89\textwidth]{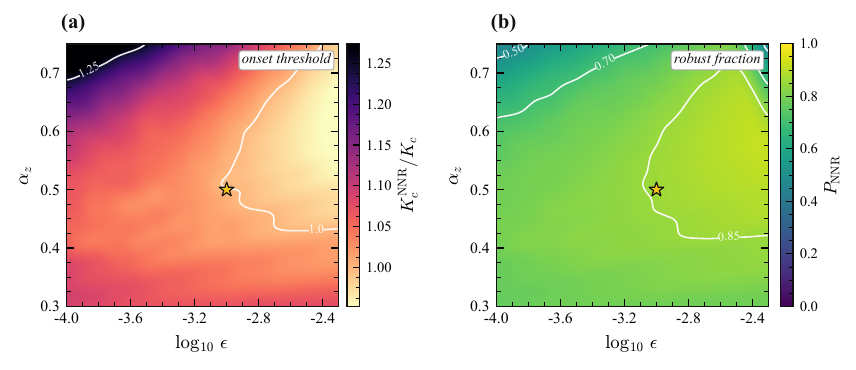} 
\caption{ For each pair \((\alpha_z,\varepsilon)\), the non-normality strength \(K\) is scanned and clean non-normal chaos is identified by the joint conditions
\(\lambda_1>0\) and \(\rho_{\max}<1\). (a) Normalized onset threshold \(K_c^{\rm NNR}(\alpha_z,\varepsilon)/K_c\), where \(K_c^{\rm NNR}\) (\ref{eq:Kc_NNR_robust}) is the smallest value of \(K\) satisfying the joint conditions ($\lambda_1>0$ and $\rho_{\max}<1$). $K_c=1.953$ is the value obtained for the baseline parameter set used for the feedback-driven class shown in figure \ref{fig:classIII_nnsrt}  $(\alpha_z=0.5, \varepsilon=10^{-3}, \Omega=\pi/2)$, shown as star marks. (b) Robust fraction \(P_{\rm NNR}(\alpha_z,\varepsilon)\) (\ref{eq:P_NNR}), defined as the fraction of scanned \(K\)-values satisfying the same joint conditions. } 
\label{fig:volume_robustness} 
\end{figure*}

The two projections in Fig.~\ref{fig:volume_robustness} give consistent evidence for parameter robustness. Panel (a) shows 
the values of the normalized threshold \(K_c^{\rm NNR}/K_c\) 
where $K_c=1.953$ is the value obtained for the baseline parameter set used for the feedback-driven class shown in figure \ref{fig:classIII_nnsrt}  $(\alpha_z=0.5, \varepsilon=10^{-3}, \Omega=\pi/2)$. \(K_c^{\rm NNR}/K_c\) 
remains close to unity over a broad region surrounding the baseline point, indicating that the onset threshold for chaos is robust over a broad parameter region.
\(P_{\rm NNR}\) is large over the same region, showing that once the onset is crossed, the clean non-normal chaotic regime persists over an extended interval of \(K\). 

Together with the feedback, memory, and rotation sweeps in
Fig.~\ref{fig:parametric_robustness}, these results show that the
feedback-driven example is representative of a robust non-normal route to
chaos: the transition is controlled by increasing non-normality and
recurrent reorientation, while the local eigenvalue spectrum remains
strictly contracting on the attractor. Similar parameter scans for the
partition-reinjected, phase-prescribed, and affine-reinjected classes lead to
the same conclusion: the positive-Lyapunov regime with
\(\rho_{\max}<1\) persists over finite parameter regions. Thus the
phenomenon is not specific to the feedback-driven implementation, but is a
robust feature of the broader family of non-normal routes to chaos studied
here.


\section{Conclusion} \label{sec:conclusion} 

We have identified a route to deterministic chaos in low-dimensional
bounded maps whose instantaneous Jacobians remain pointwise spectrally
stable everywhere on the attractor. In this regime, the eigenvalues of each
one-step Jacobian stay inside the unit disk, giving the local spectral
impression of contraction. Nevertheless, the dynamics can have a positive
maximal Lyapunov exponent because the Jacobians are non-normal: transiently
amplified singular-vector directions are repeatedly regenerated and
reinserted into the tangent dynamics.

The four classes constructed here realize this mechanism in distinct ways.
In the partition-reinjected route, a state partition switches between two
transposed orientations of the same non-normal core. In the phase-prescribed
route, an independent mod-one phase prescribes the sequence of orientations.
In the feedback-driven route, the state itself generates the reorientation
through an endogenous variable. In the affine-reinjected route, boundedness
and recurrence are achieved by branch-dependent affine translations rather
than by modulo-one wrapping. Across all four cases, the local spectrum
remains subcritical while the product of Jacobians becomes unstable:
\[
    \rho_{\mathrm{traj}}^{\max}<1,
    \qquad
    \sigma_{\mathrm{traj}}^{\max}>1,
    \qquad
    \lambda_1>0
\]
above a class-dependent transition occurring as a function of a control parameter 
$K$ (\ref{eq:K_def}) quantifying the strength of non-normality of the shared core matrix. 
Increasing this parameter changes the eigenvector geometry while
leaving the eigenvalues fixed. The matrix-level threshold for singular-value
amplification indicates when transient growth becomes possible, whereas the
actual Lyapunov threshold depends on how the dynamics regenerates and
reinjects the amplified directions. This separates local transient
amplification from sustained chaotic instability.

This mechanism is closely related to classical product effects in
nonautonomous linear systems, including the Perron phenomenon
\cite{Perron1930}, where instantaneous stability does not guarantee
stability of the evolution operator. More generally, the asymptotic
stability of time-dependent linear systems is governed by the evolution
operator, or by Lyapunov and Sacker--Sell spectra, rather than by the
instantaneous eigenvalues alone \cite{Coppel1978,SackerSell1978}. The same
perspective also connects the present construction to transient growth in
linearly stable shear flows \cite{Grossmann1994,Schmid2007}, to switched
and hybrid systems in which products of individually stable matrices can be
unstable \cite{Liberzon2003,DeCarlo2000}, and to products of random
matrices, where asymptotic growth is controlled by products rather than by
single-step spectra \cite{Crisanti1993}.

The contribution of the present work is to realize this product-growth
principle in autonomous, low-dimensional, bounded nonlinear maps and to show
that it can produce sustained deterministic chaos without any local
eigenvalue instability. These results refine the usual association between
chaos and local spectral expansion. Eigenvalues remain useful for diagnosing
pointwise contraction, but they do not bound finite-time growth when the
Jacobians are non-normal. The relevant instability is an instability of
products: singular directions are transiently amplified, and the dynamics
repeatedly restores their alignment. Thus non-normality, together with
recurrent reinjection, provides an organizing mechanism for deterministic
chaos that is distinct from classical eigenvalue-based routes.


\section*{Appendix: Methodology}
\label{sec:supp_methods}

All numerical diagnostics were generated using Python scripts that sweep the
non-normality parameter \(\kappa\), convert it to the common index
\[
    K=\frac{\kappa-\kappa^{-1}}{2},
\]
and then normalize by the class-specific Lyapunov threshold \(K_c^{(j)}\).
For each class \(j\), \(K_c^{(j)}\) was estimated from the first zero crossing
of the maximal Lyapunov exponent,
\[
    \lambda_1^{(j)}(K_c^{(j)})=0,
\]
using linear interpolation of the computed \(\lambda_1(K)\) curve.  The
horizontal axes in the main figures are therefore \(K/K_c^{(j)}\), not
\(\kappa/\kappa_c\).

For each parameter value, we computed dynamical and geometric diagnostics
from long post-transient trajectories of the corresponding map
\[
    \mathbf{X}_{n+1}=F_K^{(j)}(\mathbf{X}_n),
\]
where \(\mathbf{X}_n\) denotes the state of the class under consideration.
For Classes I and IV, \(\mathbf{X}_n\in\mathbb{R}^2\) or
\(\mathbb{T}^2\).  For Classes II and III,
\(\mathbf{X}_n=(x_n,y_n,z_n)\) includes the orientation or switching
variable.  Unless stated otherwise, the common non-normal core matrix uses
\[
    \alpha=0.7,\qquad \beta=0.2,
\]
so that its eigenvalues are fixed at
\[
    \lambda_\pm=\alpha\pm\beta,
    \qquad
    \rho(A_0)=0.9<1.
\]
The remaining parameters are class-specific and are reported in the
corresponding figure captions and source scripts.

\subsection{Lyapunov exponents}
\label{subsec:supp_lyapunov}

Let
\[
    J_n = DF_K^{(j)}(\mathbf{X}_n)
\]
denote the Jacobian matrix evaluated along a trajectory of class \(j\).
The \(N\)-step tangent propagator is
\[
    M_N(\mathbf{X}_0)=J_{N-1}J_{N-2}\cdots J_0.
\]
The Lyapunov exponents are the asymptotic logarithmic growth rates of the
singular values of \(M_N\),
\[
    \lambda_i
    =
    \lim_{N\to\infty}
    \frac{1}{N}
    \log \sigma_i\!\left(M_N(\mathbf{X}_0)\right),
    \qquad
    \lambda_1\geq\lambda_2\geq\cdots,
\]
for typical initial conditions on the invariant set
\cite{Benettin1980Numerical,benettin1980lyapunov}.  We used the standard
Benettin--QR procedure.  After discarding an initial transient, an
orthonormal tangent basis \(Q_0\) was initialized and iterated according to
\[
    Z_n=J_nQ_n,
    \qquad
    Z_n=Q_{n+1}R_n,
\]
where \(Q_{n+1}\) is orthonormal and \(R_n\) is upper triangular.  The
finite-time Lyapunov exponents were estimated as
\[
    \widehat{\lambda}_i(N)
    =
    \frac{1}{N}
    \sum_{n=0}^{N-1}\log |(R_n)_{ii}|.
\]
The number of tangent vectors was chosen to match the dimension of the class:
two for the planar maps and three for the skew-product maps.  The quantity
reported in the main text is the largest exponent \(\lambda_1\).  The sign
of \(\lambda_1\) determines the transition used to define \(K_c^{(j)}\):
\(\lambda_1<0\) indicates net asymptotic contraction, while
\(\lambda_1>0\) signals sensitive dependence on initial conditions.

\subsection{Spectral-radius and singular-value diagnostics}
\label{subsec:supp_spectral_singular}

For each sampled trajectory, we also computed one-step Jacobian diagnostics.
The instantaneous spectral radius is
\[
    \rho_n=\rho(J_n),
\]
and the instantaneous largest singular value is
\[
    \sigma_n=\sigma_{\max}(J_n)=\|J_n\|_2.
\]
The figures report these quantities in logarithmic form as \del{\(\ln\rho\) and
\(\ln\sigma\),}  \(\ln \rho_{\mathrm{traj}}^{\max}\) and
\(\ln \sigma_{\mathrm{traj}}^{\max}\), using either trajectory averages or sampled maxima depending
on the class and diagnostic file.  The central spectral-stability check is
that the sampled spectral radius remains below unity in the reported
non-normal chaotic regime,
\[
    \rho_{\mathrm{traj}}^{\max}<1.
\]
The singular-value diagnostic is used separately to quantify transient
non-normal amplification.  Thus the reported transition is not identified
from \(\sigma_{\max}>1\) alone.  The chaos threshold \(K_c^{(j)}\) is always
defined from the Lyapunov crossing \(\lambda_1^{(j)}=0\).

\subsection{Kaplan--Yorke dimension}
\label{subsec:supp_KY}

From the full Lyapunov spectrum, we computed the Kaplan--Yorke dimension
\[
    D_{\mathrm{KY}}
    =
    j+\frac{\lambda_1+\cdots+\lambda_j}{|\lambda_{j+1}|},
\]
where \(j\) is the largest integer such that
\[
    \sum_{i=1}^{j}\lambda_i \geq 0,
    \qquad
    \sum_{i=1}^{j+1}\lambda_i < 0
\]
\cite{KaplanYorke1979,Frederickson1983}.  This is a spectrum-based estimate
of attractor dimension.  It is not identical to a box-counting or
correlation dimension, but it provides a complementary measure of the
transition from periodic or low-complexity dynamics to a strange attractor.

\subsection{Occupied-bin box-counting estimates}
\label{subsec:supp_boxcount}

To quantify the sampled attractor geometry directly, we estimated
finite-resolution box-counting dimensions.  For a compact set \(A\), the
box-counting dimension is
\[
    D_0(A)
    =
    \lim_{\varepsilon\to 0}
    \frac{\log N(\varepsilon)}{\log(1/\varepsilon)},
\]
where \(N(\varepsilon)\) is the minimum number of boxes of side length
\(\varepsilon\) required to cover \(A\) \cite{Falconer2014}.  Numerically,
we used uniform grids.  For a grid with \(m\) bins per coordinate direction,
\[
    \varepsilon=\frac{1}{m}.
\]
For a sampled point cloud, the occupied-bin count \(N(m)\) was computed over
a prescribed set of grid resolutions, and the finite-resolution estimate was
obtained as the least-squares slope
\[
    D_0^{(\mathrm{est})}
    =
    \mathrm{slope}\left[
        \log N(m)\ \text{vs.}\ \log m
    \right].
\]

For planar maps or planar projections we report BC\((xy)\).  For
three-dimensional skew-product maps we also report BC\((xyz)\).  These
curves are finite-sample geometric proxies, not exact asymptotic dimensions.
Their values depend on trajectory length, sampling density, and the selected
scaling range.  They are used here as consistency checks on the Lyapunov
transition rather than as definitive fractal-dimension estimates.

\subsection{Correlation dimension}
\label{subsec:supp_corrdim}

For Classes II and III, where the attractor is naturally represented in
\((x,y,z)\), we additionally computed a correlation-dimension estimate
\(D_2\) using a Grassberger--Procaccia-type procedure
\cite{GrassbergerProcaccia1983PRL,GrassbergerProcaccia1983PhysicaD}.  For a
point cloud \(\{\mathbf{X}_i\}_{i=1}^{M}\subset\mathbb{R}^d\), the
correlation sum is
\[
    C(r)
    =
    \frac{2}{M(M-1)}
    \sum_{1\leq i<j\leq M}
    \Theta\!\left(r-\|\mathbf{X}_i-\mathbf{X}_j\|\right),
\]
where \(\Theta\) is the Heaviside function.  The correlation dimension is
defined by
\[
    D_2
    =
    \lim_{r\to 0}
    \frac{d\log C(r)}{d\log r}.
\]
In practice, \(D_2\) was estimated as the least-squares slope of
\(\log C(r)\) versus \(\log r\) over an intermediate scaling window.  The
smallest radii were excluded to reduce undersampling effects, while the
largest radii were excluded to avoid saturation.  Temporal redundancy was
reduced by subsampling from longer post-transient trajectories.  The
resulting \(D_2\) should be interpreted as a corroborative finite-sample
dimension estimate.

\subsection{Bifurcation projections}
\label{subsec:supp_bifurcation}

To generate the bifurcation-style projections, we simulated one long
post-transient orbit for each scanned value of \(K/K_c^{(j)}\).  After
discarding a burn-in, a fixed number of iterates was recorded and the
coordinate \(x_n\) was plotted against \(K/K_c^{(j)}\).  For two-dimensional
maps, \(x_n\) is the first coordinate of the planar state.  For the
skew-product maps, \(x_n\) is the first coordinate of the planar component
\((x_n,y_n)\).  These projections are not used to define the transition.
They provide a visual check that the Lyapunov crossing is accompanied by a
change from a sparse or low-complexity orbit to a broadened attracting set.

    \bibliography{bibliography} 

\end{document}